# Performance analysis of hybrid FSO/RF communication systems with Alamouti Coding or Antenna Selection

*Javad Sayevand[1], Mohammad Ali Amirabadi[1]*
[1] *School of Electrical Engineering, Iran University of Science and Technology, Tehran, Iran*
E-mail: m_amirabadi@elec.iust.ac.ir

**Abstract:** In this work a novel dual-hop relay-assisted hybrid Free Space Optical / Radio Frequency (FSO / RF) communication system is presented. In this system, RF signal is transmitted from two antennas, and then forwarded by a single antenna relay through FSO channel. This is the first time that performance of using Alamouti Coding (AC) or Antenna Selection (AS) at the transmitter of a hybrid FSO / RF system is investigated. FSO link has Gamma-Gamma atmospheric turbulence, and in order to get closer to the actual results, the effect of pointing error is also considered. For the first time closed-form expressions are derived for Bit Error Rate (BER) and Outage Probability ($P_{out}$) of the proposed system and validated through MATLAB simulations. Results indicate that in this structure, there is slight performance difference between AC and AS schemes. Hence due to more complexity, power consumption and latency of AC, AS is recommended. Dual-hop, hybrid FSO / RF system significantly improves performance and reliability of the system, and is particularly suitable for long-range applications that direct RF communication between source and destination is not possible. Considering these advantages this structure is particularly suitable for mobile communications which has power and processing limitations.

## 1 Introduction

In FSO communication systems, often Intensity Modulation / Direct Detection (IM / DD) based on on-off keying (OOK), is used due to its simplicity. In OOK, detection threshold is adjusted based on atmospheric turbulence intensity; accordingly, it is suitable for areas with varying turbulence intensity. Pulse Position Modulation (PPM) is another modulation used in FSO system, which does not need adaptive detection threshold. Subcarrier Intensity Modulation (SIM) does not require adaptive detection threshold and compared with PPM has higher spectral efficiency [1].

Space-Time Block Coding (STBC), is one of the techniques used in wireless communication systems. STBC collects various copies of transmitted data with the help of multiple antennas. Correct combination of these copies at the receiver will significantly improve performance of the system. Among various STBCs, Alamouti Coding (AC) is the easiest orthogonal STBC used for Multiple Input Multiple Output systems and has full diversity and unit code rate [1].

Bandwidth and license constraints of RF systems, which offer data rate up to 100Mbps, encouraged communication companies to find a better solution for the need of bandwidth and data rate. FSO system which provides up to 2000 THz unlicensed bandwidth is an appropriate solution [2]. However, beside FSO advantages, its sensitivity to weather conditions and atmospheric turbulences severely limits its practical implementation [3].

An efficient way to improve performance of RF link is to combine it with a FSO link. These systems which are called hybrid FSO / RF, are available in series [4] or parallel [5] structures. In series structure, data is transmitted via RF or FSO link, and then forwarded by a relay through another FSO or RF link. Use of a relay improves performance and capacity of the system [6]. In parallel structures, RF and FSO links are parallel and send data simultaneously [7] or by use of a switch [8].

The main difference between relay assisted communication systems is related to the processing they make on the signal. Amplify and Forward (AF) relaying is mostly used due to its simple structure and easy implementation [9-11]. The amplification gain can be fixed [12] or variable based on channel conditions [13]. In AF noise is also amplified significantly, thereby some studies use other schemes such as Detect and Forward (DF) [14,15] and Quantize and Forward (QF) [16]. In DF received signal is detected and Forwarded by the relay. In QF, log-likelihood of data symbol is estimated and quantized by the relay.

This work presents a simple novel idea for combination of space diversity and relaying scheme. To the best of the authors' knowledge there is no other work in hybrid FSO/RF dealing with transmit diversity and Other papers published on FSO / RF just focused on different atmospheric turbulences or relaying protocols; The authors, in this paper and some other papers [17-20] have tried to combine various space diversity schemes with hybrid FSO/RF system. Authors thought that it might be interesting to show what happens when these two techniques are combined. So they have considered a complex and a simple diversity scheme, i.e. Alamouti Coding which is the complex because it's taken from Maximum Ratio Combiner (MRC), and Antenna Selection which is simple because just compares the transmitted signals.

In this paper two novel transmission schemes are considered for hybrid FSO / RF system. Transmit diversity greatly improves system performance. AC and Antenna Selection (AS) are two transmit diversity schemes that are widely used in RF systems. However it is the first time that the effect of AC or AS on the performance of hybrid FSO / RF systems is investigated. IM / DD is used in FSO link. FSO link is considered at Gamma-Gamma [21] atmospheric turbulence because this model is highly accompanied with experimental results. Also in order to get closer to actual operations, the effect of pointing error [22] is considered at FSO link. RF link is at Rayleigh fading. For the first time, closed-form expressions are derived for BER and $P_{out}$ of the proposed structure and validated via MATLAB simulation. Dual-hop, hybrid FSO / RF system significantly improves performance and link reliability, and is particularly suitable for long-range, also it reduces total power consumption, and increases capacity.

Rest of this paper is organized as follows: Section II describes system model of AC scheme and evaluates its performance. Section

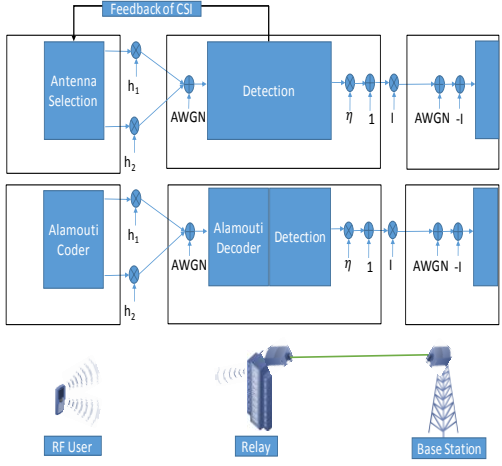

**Fig. 1**: Proposed hybrid FSO / RF structure.

III discusses system model and evaluates performance of AS scheme. Section IV, provides simulations and discussions and section V is the conclusion of this study.

## 2 Alamouti Coding Scheme

### 2.1 System Model

In Fig. 1, the proposed dual hop relay assisted hybrid FSO/RF system is considered in which a relay is connected to mobile user via RF link and to base station via FSO link. Mobile user uses Alamouti Coder for data transmission. Relay uses Alamouti Decoder and converts detected RF signal to FSO signal with conversion efficiency of $\eta$, and then a DC bias with unit amplitude is added to FSO signal in order to FSO signal be positive. In this system $x_1, x_2$ are the transmitted symbols in two consecutive time slots in the form of Alamouti code [23]:

$$X = \begin{bmatrix} x_1 & -x_2^* \\ x_2 & x_1^* \end{bmatrix}, \quad (1)$$

where $(.)^*$ is complex conjugate. These transmitted signals pass through Rayleigh fading channel. Received signals at the relay, in two consecutive time slots, are as:

$$[y_{1,1} \quad y_{1,2}]^T = [x_1 h_1 + x_2 h_2 \quad -x_2^* h_1 + x_1^* h_2]^T + [e_{1,1} \quad e_{1,2}]^T, \quad (2)$$

where $e_{1,1}, e_{1,2}$ are Additive White Gaussian Noise (AWGN) with zero mean and $\sigma_1^2$ variance, at the relay input in the first and second time slots, $h_i, i = 1.2$ is fading coefficient of RF link between relay and $i^{th}$ transmit antenna, and $[.]^T$ is transposed. After conjugating the received signal in second time slot ($y_{1,2}$), and multiplying $\begin{bmatrix} h_1^* & h_2 \\ h_2^* & -h_1 \end{bmatrix}$ from the left side, (2) becomes:

$$[r_{1,1}, r_{1,2}]^T = \begin{bmatrix} y_{1,1} h_1^* + y_{1,2}^* h_2 \\ y_{1,1} h_2^* - y_{1,2}^* h_1 \end{bmatrix} = \begin{bmatrix} (|h_1|^2 + |h_2|^2) x_1 + n_{1,1} \\ (|h_1|^2 + |h_2|^2) x_2 + n_{1,2} \end{bmatrix}, \quad (3)$$

where $n_{1,1} = e_{1,1} h_1^* + e_{1,2}^* h_2$ and $n_{1,2} = e_{1,1} h_2^* + e_{1,2}^* h_1$ are AWGN with zero mean and $\sigma_1^2(|h_1|^2 + |h_2|^2)$ variance. Accordingly, if $x$ is transmitted in a time slot the received signal at the relay will be in the following form:

$$r_1 = (|h_1|^2 + |h_2|^2)x + n_1 = |h|^2 x + n_1, \quad (4)$$

where $n_1$ is AWGN with zero mean and $\sigma_1^2(|h_1|^2 + |h_2|^2)$ variance. The relay detects the received RF signal and converts it to FSO with conversion efficiency $\eta$, then adds a unit amplitude DC bias to it. The forwarded FSO signal is as follows:

$$x_R = 1 + \eta d_1, \quad (5)$$

where $d_1$ is the detected signal. After DC removal, the received FSO signal becomes as follows:

$$r_2 = I x_R + n_2 - I = I \eta d_1 + n_2, \quad (6)$$

where $I$ is FSO link atmospheric turbulence intensity and $n_2$, is AWGN with zero mean and $\sigma_2^2$ variance, at the FSO receiver input.

### 2.2 Outage Probability

In proposed structure, the outage of total system is caused by the outage of relay or Base Station. Assuming independent detections at relay and destination, $P_{out}$ of the proposed structure becomes as follows [8]:

$$P_{out}(\gamma_{th}) = Pr\left\{\cup_{i=1}^2 \{i^{th} \text{ link is in outage}\}\right\} = 1 - Pr\left\{\cap_{i=1}^2 \{i^{th} \text{ link is available}\}\right\} = 1 - [1 - P_{out,RF}(\gamma_{th})][1 - P_{out,FSO}(\gamma_{th})]. \quad (7)$$

CDF of Gamma-Gamma atmospheric turbulence with the effect of pointing error is as follows [24]:

$$F_\gamma(\gamma) = \frac{\xi^2}{\Gamma(\alpha)\Gamma(\beta)} G_{2,4}^{3,1}\left(\alpha\beta\kappa\sqrt{\frac{\gamma}{\bar{\gamma}_{FSO}}} \Big| \begin{matrix} 1, 1+\xi^2 \\ \xi^2, \alpha, \beta, 0 \end{matrix}\right), \quad (8)$$

where $G_{\cdot,\cdot}^{\cdot,\cdot}\left(-\Big|\begin{matrix} - \\ - \end{matrix}\right)$ and $\Gamma(.)$ are the well-known Meijer-G and Gamma functions, respectively. $\alpha, \beta$ are small scale and large scale characteristic parameters of Gamma-Gamma atmospheric turbulence and $\xi^2$ is characteristic parameter of pointing error. $\bar{\gamma}_{FSO} = \eta^2 E[x^2]/\sigma_2^2$ is average SNR at the Base Station input. The CDF of Rayleigh fading is as follows:

$$F_\gamma(\gamma) = 1 - e^{-\frac{\gamma}{\bar{\gamma}_{RF}}}. \quad (9)$$

From (4), instantaneous SNR at relay the input is equal to:

$$\gamma_{RF} = \frac{(|h_1|^2 + |h_2|^2)E[x^2]}{\sigma_1^2} = \bar{\gamma}_{RF}|h_1|^2 + \bar{\gamma}_{RF}|h_2|^2 = \gamma_1 + \gamma_2. \quad (10)$$

where $\bar{\gamma}_{RF} = E[x^2]/\sigma_1^2$ is average SNR at the relay input. In this paper, the Moment Generating Function (MGF) approach is used to obtain CDF of RF link. Given that the MGF is Laplace transform of the pdf, and assuming independent and identically distributed RF paths, and using (10), MGF of RF link is as follows:

$$M_{\gamma_{RF}}(s) = M_{\gamma_1}(s) M_{\gamma_2}(s) = \left(\frac{1}{s\bar{\gamma}_{RF}+1}\right)^2. \quad (11)$$

Taking inverse Laplace transform of (11), and then integrating, the CDF of RF link becomes equal to:

$$F_{\gamma_{RF}}(\gamma) = 1 - \left(1 + \frac{\gamma}{\bar{\gamma}_{RF}}\right) e^{-\frac{\gamma}{\bar{\gamma}_{RF}}}. \quad (12)$$

Substituting (8) and (12) in (7), $P_{out}$ of the proposed system is equal to:

$$P_{out}(\gamma_{th}) = 1 - \left(1 + \frac{\gamma_{th}}{\bar{\gamma}_{RF}}\right) e^{-\frac{\gamma_{th}}{\bar{\gamma}_{RF}}} \Big[ 1 - \frac{\xi^2}{\Gamma(\alpha)\Gamma(\beta)} G_{2,4}^{3,1}\left(\alpha\beta\kappa \sqrt{\frac{\gamma_{th}}{\bar{\gamma}_{FSO}}} \Big| \begin{matrix} 1,1+\xi^2 \\ \xi^2, \alpha, \beta, 0 \end{matrix} \right) \Big]. \quad (13)$$

## 2.3 Bit Error Rate

In coherent modulation, phase recovery error degrades system performance, while differential modulation schemes such as DPSK are less sensitive to it. Receiver with incoherent detection is also less complex. Given that $F_\gamma(\gamma) = P_{out}(\gamma)$, BER of DPSK modulation is obtained from the following equation [4]:

$$P_e = \frac{1}{2} \int_0^\infty e^{-\gamma} F_\gamma(\gamma)\, d\gamma = \frac{1}{2} \int_0^\infty e^{-\gamma} P_{out}(\gamma)\, d\gamma. \quad (14)$$

Substituting (13) in (14) and using [25], BER of DPSK modulation becomes as follows:

$$P_e = \frac{1}{2} - \frac{\left(1 + \frac{2}{\bar{\gamma}_{RF}}\right)}{2\left(1 + \frac{1}{\bar{\gamma}_{RF}}\right)^2} + \frac{\xi^2 2^{\alpha+\beta-4}}{\pi \Gamma(\alpha)\Gamma(\beta)} \Bigg[ \frac{1}{\left(1 + \frac{1}{\bar{\gamma}_{RF}}\right)} G_{5,8}^{6,3}\left(\frac{(\alpha\beta\kappa)^2}{16\bar{\gamma}_{FSO}\left(1+\frac{1}{\bar{\gamma}_{RF}}\right)} \Big| \begin{matrix} \psi_1 \\ \psi_2 \end{matrix}\right) + \frac{1}{\bar{\gamma}_{RF}} \frac{1}{2\left(1 + \frac{1}{\bar{\gamma}_{RF}}\right)^2} G_{5,8}^{6,3}\left(\frac{(\alpha\beta\kappa)^2}{16\bar{\gamma}_{FSO}\left(1+\frac{1}{\bar{\gamma}_{RF}}\right)} \Big| \begin{matrix} \psi_3 \\ \psi_2 \end{matrix}\right) \Bigg], \quad (15)$$

where $\psi_1 = \left(0, \frac{1}{2}, 1, \frac{1+\xi^2}{2}, \frac{2+\xi^2}{2}\right)$, $\psi_2 = \left(\frac{\xi^2}{2}, \frac{\xi^2+1}{2}, \frac{\alpha}{2}, \frac{\alpha+1}{2}, \frac{\beta}{2}, \frac{\beta+1}{2}, 0, \frac{1}{2}\right)$, and $\psi_3 = \left(-1, \frac{1}{2}, 1, \frac{1+\xi^2}{2}, \frac{2+\xi^2}{2}\right)$.

## 3 Antenna Selection Scheme

### 2.4 System Model

As shown in Fig. 1, in AS scheme, mobile user selects better antenna according to the received feedback from relay. Then relay converts detected RF signal to FSO signal with conversion efficiency of $\eta$, and a unit amplitude DC bias is added to FSO signal to be positive. AS selects the antenna that its signal has maximum SNR at the relay input, i.e.

$$\gamma_{RF} = max(\gamma_{1,1}, \gamma_{1,2}), \quad (16)$$

where, $\gamma_{1,i}$ is instantaneous SNR at the relay input from $i^{th}$ transmit antenna. Accordingly, and assuming $x$ as the transmitted signal at a time slot, the received signal at the relay will be in the following form:

$$r_1 = hx + n_1 \quad (17)$$

Assuming independent and identically distributed RF paths, and using (9) and (16), the CDF of RF link becomes as follows:

$$F_{\gamma_{RF}}(\gamma) = Pr(max(\gamma_{1,1}, \gamma_{1,2}) \leq \gamma) = Pr(\gamma_{1,1} \leq \gamma, \gamma_{1,2} \leq \gamma) = F_{\gamma_{1,i}}(\gamma) F_{\gamma_{2,i}}(\gamma) = \left(1 - e^{-\frac{\gamma}{\bar{\gamma}_{RF}}}\right)^2. \quad (18)$$

### 2.5 Outage Probability

Given that $P_{out}(\gamma_{th}) = F_\gamma(\gamma_{th})$, substituting (8) and (18) into (7), $P_{out}$ of the proposed hybrid FSO / RF system becomes:

$$P_{out}(\gamma_{th}) = 1 - 2e^{-\frac{\gamma_{th}}{\bar{\gamma}_{RF}}} + e^{-\frac{2\gamma_{th}}{\bar{\gamma}_{RF}}} + \frac{\xi^2}{\Gamma(\alpha)\Gamma(\beta)} \left(2e^{-\frac{\gamma_{th}}{\bar{\gamma}_{RF}}} - e^{-\frac{2\gamma_{th}}{\bar{\gamma}_{RF}}}\right) G_{2,4}^{3,1}\left(\alpha\beta\kappa \sqrt{\frac{\gamma_{th}}{\bar{\gamma}_{FSO}}} \Big| \begin{matrix} 1,1+\xi^2 \\ \xi^2, \alpha, \beta, 0 \end{matrix}\right). \quad (19)$$

### 2.6 Bit Error Rate

Given that $P_{out}(\gamma) = F_\gamma(\gamma)$ and substituting (19) into (14), and using [25] BER of the proposed structure becomes:

$$P_e = \frac{1}{2} - \frac{1}{1 + \frac{1}{\bar{\gamma}_{RF}}} + \frac{\frac{1}{2}}{1 + \frac{2}{\bar{\gamma}_{RF}}} + \frac{\xi^2 2^{\alpha+\beta-3}}{\pi \Gamma(\alpha)\Gamma(\beta)} \Bigg( \frac{1}{\left(1 + \frac{1}{\bar{\gamma}_{RF}}\right)} G_{5,8}^{6,3}\left(\frac{(\alpha\beta\kappa)^2}{16\bar{\gamma}_{FSO}\left(1+\frac{1}{\bar{\gamma}_{RF}}\right)} \Big| \begin{matrix} \psi_1 \\ \psi_2 \end{matrix}\right) - \frac{\frac{1}{2}}{\left(1 + \frac{2}{\bar{\gamma}_{RF}}\right)} G_{5,8}^{6,3}\left(\frac{(\alpha\beta\kappa)^2}{16\bar{\gamma}_{FSO}\left(1+\frac{2}{\bar{\gamma}_{RF}}\right)} \Big| \begin{matrix} \psi_1 \\ \psi_2 \end{matrix}\right) \Bigg). \quad (20)$$

## 4 Comparison of Simulation and Analytical Results

In this section, analytical and MATLAB simulation results of performance evaluation of the proposed structure are compared. FSO link has moderate ($\alpha = 4, \beta = 1.9, \xi = 10.45$) and strong ($\alpha = 4.2, \beta = 1.4, \xi = 2.45$) regimes of Gamma-Gamma atmospheric turbulence with the effect of pointing error; RF link has Rayleigh fading. Average SNR of FSO and RF links are assumed to be equal ($\gamma_{avg} = \bar{\gamma}_{FSO} = \bar{\gamma}_{RF}$), also it is assumed that $\eta = 1$. $\gamma_{th}$ is Outage threshold SNR. In MATLAB simulations Gamma-Gamma random variable is generated by multiplication of two Gamma random variables using gamrnd(.) command. Also for radial displacement of pointing error, Rayleigh model is considered which is complex addition of two Gaussian random variables using randn(.) command.

In Fig. 2, plots Bit Error Rate of AC and AS schemes in terms of average SNR for moderate ($\alpha = 4, \beta = 1.9, \xi = 10.45$) and strong ($\alpha = 4.2, \beta = 1.4, \xi = 2.45$) regimes of Gamma-Gamma atmospheric turbulence with the effect of pointing error. As can be seen, in the proposed structure performances of AC and AS schemes are very close. Although AC and AS, both require channel state information, but AS has lower power consumption and processing delay, and is cost effective. At $\gamma_{avg} \leq 15$dB, there is slight performance difference between moderate and saturate atmospheric turbulence regimes. This is because at low $\gamma_{avg}$, the effect of the noise is dominant and at high $\gamma_{avg}$ the effect of atmospheric turbulence is dominant. These two effects have different physical phenomena. The proposed structure does not require additional processing to adaptively adjust its parameters, also it does not require much more power consumption to maintain performance. Therefore, cost and power consumption of the proposed system are effective. The proposed structure is particularly suitable for communication systems with constraints on power consumption, complexity.

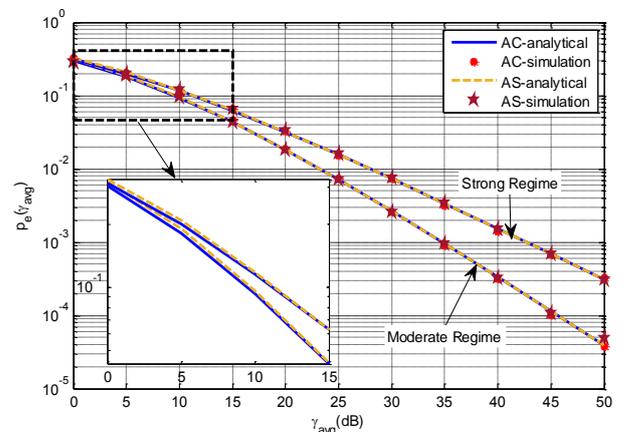

**Fig. 2**: Bit Error Rate of AC and AS schemes in terms of average SNR for moderate ($\alpha = 4, \beta = 1.9, \xi = 10.45$) and strong ($\alpha = 4.2, \beta = 1.4, \xi = 2.45$) regimes of Gamma-Gamma atmospheric turbulence with the effect of pointing error.

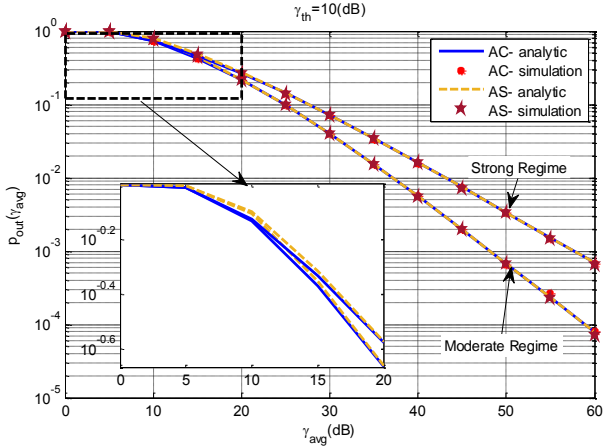

**Fig. 3**: Outage Probability of AC and SC schemes in terms of average SNR for moderate ($\alpha = 4, \beta = 1.9, \xi = 10.45$) and strong ($\alpha = 4.2, \beta = 1.4, \xi = 2.45$) regimes of Gamma-Gamma atmospheric turbulence with pointing errors when $\gamma_{th} = 10dB$.

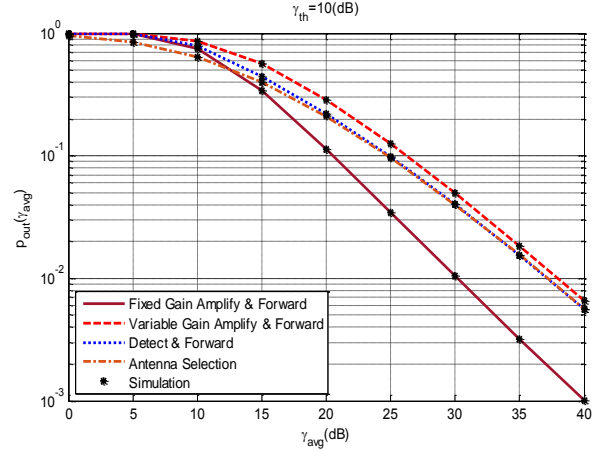

**Fig. 4**: Outage Probability in terms of average SNR for moderate ($\alpha = 4, \beta = 1.9, \xi = 10.45$) regime of Gamma-Gamma atmospheric turbulence with the effect of pointing error.

The application scenario of AS scheme is anywhere that requires low power and low complexity communication such as mobile communications in which the transmitter is a mobile with a simple processing and small battery.

Base Stations usually use AC in spite of its complexity and complexity which brings more power consumption and latency; then the question is how could solve these problems while maintaining performance as well as structure? This paper answers my question; in these situations and other situations like that, it is very better to use the proposed AS scheme.

The proposed structures are timely because in this age there are huge demands for communication speed; as well, electronic devices such as mobiles need to supply their power more while maintaining performance; so they require less processing to do that.

If combine two systems could gather their advantages at once. Therefore it's more practical and useful than considering just one of them. So this work is more practical than exciting FSO/RF works.

In Fig. 3, plots Outage Probability of AC and AS schemes in terms of average SNR for moderate ($\alpha = 4, \beta = 1.9, \xi = 10.45$) and strong ($\alpha = 4.2, \beta = 1.4, \xi = 2.45$) regimes of Gamma-Gamma atmospheric turbulence with the effect of pointing errors, when $\gamma_{th} = 10dB$. As can be seen, AC and AS have very slight performance difference at low $\gamma_{avg}$, and at the rest $\gamma_{avg}$, they perform the same. AC has more processing delay, therefore in time sensitive applications, AS is recommended. Also in the proposed system, complexity and power consumption of AC scheme encourages the mobile operators to use AS.

In Fig.4, Outage Probability is plotted in terms of average SNR for moderate ($\alpha = 4, \beta = 1.9, \xi = 10.45$) regime of Gamma-Gamma atmospheric turbulence with the effect of pointing errors, when $\gamma_{th} = 10dB$. AC scheme is not plotted because of its similarity to AS structure. This figure compares AS scheme of the proposed structure with single user case of two most similar dual-hop structures [11], and [15], at the first structure an RF link is followed by a FSO link and its relay uses Amplify & Forward relaying; but at the second structure RF link is followed by parallel FSO/RF structure and its relay uses Detect & Forward relaying. At these structures RF signal is transmits from single antenna without any coding or diversity. It has worth to remind that the proposed structure uses Detect & Forward relaying.

As can be seen, the proposed structure performs better than variable gain Amplify & Forward at all $\gamma_{avg}$. But it has the same performance as Detect & Forward scheme at high $\gamma_{avg}$; the reason is that in Detect & Forward scheme, RF link is followed by a parallel FSO / RF link; but in the proposed structure it's followed by FSO link. FSO and RF links are complementary of each other, and there is almost no atmospheric condition which could degrade them both. This is an advantage of the proposed structure that while using FSO link at the second hop, could achieve performance of parallel FSO/RF link. It's cost and processing affordable. Also it is shown that at high $\gamma_{avg}$, fixed gain Amplify & Forward performs better. This is because that in fixed gain Amplify and Forward, the amplification gain is adjusted manually, and one can rise it as high as he wants; but it's not economically affordable and consumes much more power.

## 5 Conclusion

This work presents a simple novel idea for combination of space diversity and relaying FSO/RF. To the best of authors' knowledge there is no other work in FSO/RF dealing with transmit diversity; they thought that it could be interesting to show what happens when these two techniques are combined. So they have considered a complex and a simple diversity scheme, i.e. AC which is the complex because it's taken from MRC, and Antenna Selection which is simple because just compares the transmitted signals.

Considering FSO link in Gamma-Gamma atmospheric turbulence with the effect of pointing errors and RF link in Rayleigh fading, for the first time, closed-form expressions are derived for $P_{out}$ and BER of the proposed system and validated through MATLAB simulations. It is shown that use of AC and AS schemes at the transmitter of a dual-hop hybrid FSO / RF system have the same performances. However, use of AS, due to its lower power consumption, simplicity and lower latency, is recommended. Also it is observed that at low average SNR, the proposed system performs almost the same at moderate and strong atmospheric turbulence intensities. So use of the proposed structure is recommended for atmospheric turbulences with frequent changes, because the proposed structure does not require adaptive processing or much more power consuming in order to maintain performance. The proposed Detect & Forward structure is compared to two other structures; it's shown that the proposed structure that while using FSO link at the second hop, could achieve performance of parallel

FSO/RF link at the second hop, and therefore it's more cost and processing affordable.